\documentstyle[prl,aps,multicol,psfig]{revtex}
\begin{document}

\draft

\title{Quantum Evaporation of a Bose-Einstein Condensate}

\author{R.A. Duine and H.T.C. Stoof}
\address{Institute for Theoretical Physics,
         University of Utrecht, Leuvenlaan 4, \\
         3584 CE Utrecht, The Netherlands}

\maketitle

\begin{abstract}
We show that a Bose-Einstein condensate emits atoms, if either the
condensate wave function, or the scattering length of the atoms depends
strongly on time. Moreover, the emission process is coherent and atoms
can oscillate back and forth between the condensate and the excited
states. Inspired by recent experimental results, we present results of
simulations of the response of a Bose-Einstein condensate to a very rapid
change in the scattering length. The possibility of molecule formation is
also discussed.
\end{abstract}


% Definitions
\def\bx{{\bf x}}
\def\bk{{\bf k}}
\def\bn{{\bf n}}
\def\bm{{\bf m}}
\def\half{\frac{1}{2}}
\def\args{(\bx,t)}

\begin{multicols}{2} {\it Introduction.} --- There are several examples
where the coupling of a classical field to a quantum field leads to the
production of quanta of the latter. One of the best known examples is
Hawking radiation, where the curved spacetime provides the classical
`background' that couples to a quantum field. Another example is
bremsstrahlung, where an electron scatters with a nucleus, and emits a
quantum of the electromagnetic field, i.e., a photon. From a quantum
mechanical point of view, the emission of radiation by an antenna is also
described by the coupling between a classical current, and the photon
field. 

An example of a classical, albeit complex field, which is subject to a
high level of experimental control, is the macroscopic wave function of a
Bose-Einstein condensate. Recently, this experimental control could even
be extended to the interactions between the atoms
\cite{inouye,courteille,simon}, which are at the temperatures and
densities of interest solely determined by the $s$-wave scattering length.
Making use of a so-called Feshbach resonance \cite{eite}, one is able to
vary the scattering length of the atoms to any possible value, by tuning
the magnetic bias field. This experimental degree of freedom was recently
exploited to study the condensate collapse \cite{elizabeth}, which was
first observed in the experiment by Bradley {\it et al.} \cite{curtis}. 
Roughly speaking, such a collapse occurs if the condensate contains so
many atoms that the mean-field interactions exceed the energy-level
splitting of the external trapping potential \cite{collapse1,collapse2}.
Clearly, the wave function of a collapsing condensate is an example of a
time-dependent classical field, since it undergoes very rapid and violent
dynamics \cite{cass1,kagan,kerson2,kerson1,rembert1}.

Another type of time dependence arises if the scattering length changes on
a time scale that is fast compared to the collective modes of the
condensate, i.e., fast compared to the inverse of the frequencies of the
external trapping potential.

It is the main purpose of this Letter to point out that both kind of time
dependences, can cause the transfer of condensate atoms to excited states.
Theoretically, this is described by the coupling between a classical
field, i.e., the condensate wave function, and a quantum field which
describes the atoms in the excited states, similar to the examples
mentioned above.  Moreover, due to the fact that this quantum evaporation
process also contains a coherent part, the atoms can oscillate back and
forth between the condensate and the excited states, in a kind of
multimode Rabi oscillation. This condensate loss mechanism has therefore
the peculiar feature that at later times less atoms are lost from the
condensate. This behavior is completely different from conventional loss
mechanisms, such as dipolar relaxation and three-body recombination, which
are characterized by a rate coefficient and therefore always lead to more
atom loss after a longer time evolution. In a very recent experiment by
Claussen {\it et al.} \cite{neil}, the number of condensate atoms was
measured after a very rapid change in the scattering length. In this
experiment, it was found that the number of condensate atoms indeed
increases with time in some regimes, and therefore we believe that our
theory might offer an explanation for these experiments.

{\it Quantum evaporation.} ---  A convenient starting point for our
discussion is the second-quantized hamiltonian for the system. It is given
in  terms of the Heisenberg creation and annihilation operators, denoted
by $\hat \psi^{\dagger} (\bx,t)$ and $\hat \psi (\bx,t)$, that create and
annihilate an atom at position $\bx$ and time $t$, respectively, and obey
the usual Bose commutation relations at equal time. The hamiltonian reads
\vspace*{-0.1in}
\begin{eqnarray} 
\label{hamiltonian}
  && \hat H = \int\!d \bx
   \ \hat \psi^{\dagger} (\bx,t) {\mathcal H}_0 \hat \psi (\bx,t)  \nonumber \\ &+&
   \frac{T^{\rm 2B} (t)}{2}  \int\! d \bx \ \hat \psi^{\dagger} (\bx,t)
  \hat \psi^{\dagger} (\bx,t) \hat \psi (\bx,t) \hat \psi (\bx,t). 
\end{eqnarray}
The single-particle hamiltonian ${\mathcal H}_0$ contains the kinetic
energy of the atoms and the external trapping potential. Its eigenstates
and eigenvalues are denoted by $\chi_{\bn} (\bx)$ and $\epsilon_{\bn}$,
respectively. The strength and sign of the interactions are fully
determined by the two-body T(ransition) matrix, $T^{\rm 2B} (t) = 4 \pi a
(t) \hbar^2/m$, where $m$ is the mass of one atom. Note that the $s$-wave
scattering length $a(t)$ is explicitly allowed to depend on time, which is
experimentally realized by tuning the magnetic field near the Feshbach
resonance. 

We split the operators in a condensate part, denoted by $\phi \args$,
which will be treated as a complex classical field, and a part that
describes the fluctuations, denoted by $\hat \psi' \args$. We insert the
separation $\hat \psi = \phi +\hat  \psi'$ into the hamiltonian in
Eq.~(\ref{hamiltonian}), and keep terms up to quadratic order in the
fluctuations. This is known as the Bogoliubov approximation, and means
physically that we assume that most of the atoms are in the condensate.
This assumption is reasonable for the zero-temperature applications under
consideration here.  The hamiltonian now consists of three parts, $H_{\rm
GP}+\hat H_{\rm B}+ \hat H_{\rm int}$, where $H_{\rm GP}$ is the usual
Gross-Pitaevskii energy functional, and $\hat H_{\rm B}$ contains the
terms that are quadratic in the fluctuations. The interaction between the
condensate and the quantum fluctuations is described by $\hat H_{\rm
int}$, which, in first instance, is given by the sum of
\[
    \int\! d \bx \ \phi^* \args \Bigl[ 
    {\mathcal H}_0 
    + T^{\rm 2B} (t) |\phi \args|^2 \Bigr] \hat \psi' \args,
\]
and the hermitian conjugate expression.

Physically, we want the classical field $\phi \args$ to describe the
low-energy part of the system, i.e., the low-energy single-particle
states, whereas the fluctuations are the high-energy excited states.
Depending on the physics of the specific application, we introduce a
cut-off between these two parts of the system. This implies that we drop
the terms containing ${\mathcal H}_0$ in $\hat H_{\rm int}$, which now
takes the form of a coupling between the classical `current density' $J
\args \equiv T^{\rm 2B} (t) |\phi \args|^2 \phi \args$, and the quantum
field $\hat \psi' \args$. Because of this coupling, a perturbation of the
classical field can result in the production of quanta of the quantum
field, i.e., atoms can be transferred from the condensate to the excited
states. To study this process, and to derive a rate equation for the
number of atoms in the condensate, we have to solve the Heisenberg
equation of motion for the quantum field operators, given by
\begin{eqnarray}
\label{heisenbergeom}
  \Bigl[ i \hbar \frac{\partial}{\partial t} 
        - {\mathcal H}_0 - 2 T^{\rm 2B} (t)|\phi \args|^2
  \Bigr] \hat \psi' \args = J \args,
\end{eqnarray}
where the hermitian conjugate expression holds for the creation operator
$\hat \psi'^{\dagger} \args$. In this equation of motion, we have
neglected the anomalous parts of the hamiltonian $\hat H_{\rm B}$, since
these describe the collective motion of the condensate, and are supposed
to be included in the condensate wave function $\phi \args$. 

The Heisenberg equation of motion in Eq.~(\ref{heisenbergeom}) is most
easily solved by introducing the retarded Green's function $G^{(+)}
(\bx,t;\bx',t')$ by means of 
\begin{eqnarray}
\label{defgreens}
  \Bigl[ i \hbar \frac{\partial}{\partial t} 
        - {\mathcal H}_0 - 2 T^{\rm 2B} (t)|\phi \args|^2
  \Bigr]  G^{(+)} (\bx,t;\bx',t') \nonumber \\ = \hbar \delta (\bx-\bx') \delta
  (t-t'),
\end{eqnarray} 
with the boundary condition $G^{(+)} (\bx,t;\bx',t')=0$ for $t<t'$.
Physically, this Green's function describes the propagation of the atoms
in the excited states, in the absence of the interaction $\hat H_{\rm
int}$.  Therefore, we have that
\begin{equation}
  iG^{(+)} (\bx,t;\bx',t')\!=\!\theta (t-t') \langle [\hat \psi' \args,\hat
  \psi'^{\dagger} (\bx',t')]\rangle_{J=0}.
\end{equation}
Because of the coupling with the classical `current density' $J \args$,
the operator for the fluctuations acquires a nonzero expectation value,
given by 
\begin{eqnarray}
\label{expecvalue}
  \langle \hat \psi' \args \rangle
   = \frac{1}{\hbar} \int\! dt' \int\! d \bx' \ G^{(+)} (\bx,t;\bx',t') J
   (\bx',t').
\end{eqnarray}
Assuming that initially all the atoms are in the condensate the density of
the noncondensed atoms is given by $n' \args =| \langle \hat \psi' \args
\rangle |^2$, and the rate equation for the number of atoms $N_{\rm c}$ in
the condensate now reads
\begin{eqnarray}
\label{rateeqn}
  \frac{d N_{\rm c} (t)}{dt} &=& - \frac{d}{dt} \int\! d \bx \ n' \args \nonumber
  \\
  &=&\frac{2}{\hbar^2} \int\!d\bx\!\int\!dt'\!\int\!d\bx' 
  {\rm Im} \Bigl[
     T^{\rm 2B} (t) |\phi \args|^2 \phi^* \args 
     \nonumber \\  
    && G^{(+)} (\bx,t;\bx',t') T^{\rm 2B} (t') |\phi (\bx',t')|^2 \phi (\bx',t')
      \Bigr].
\end{eqnarray}

This equation is our most important result, and describes the change in
the number of condensate atoms due to a strong time dependence of the
condensate wave function, and/or the scattering length of the atoms. 
Since we are treating the quantum field $\hat \psi'$ as noninteracting,
the process is coherent. In particular, oscillations of atoms between the
condensate and the thermal cloud can occur.  Clearly, the rate equation is
nonmarkovian, and on short time scales there is no conservation of energy,
due to the Heisenberg uncertainty principle. However, in the markovian
limit, where the energy of the ejected atoms is taken much larger than the
energy of a condensate atom, the rate equation takes the form of Fermi's
Golden Rule for the process of an elastic collision between two condensate
atoms, where one atom is ejected from the condensate and one atom is
stimulated back into the condensate \cite{rembert1}. In particular, this
means that in equilibrium there is no correction to the usual
Gross-Pitaevskii equation. Presently, we are mainly interested in the new
features that arise due to the nonmarkovian nature of the process under
consideration. 

{\it Multimode Rabi oscillations} ---  As an example, we discuss the
response of the system to a change in the scattering length that is too
fast for the condensate to react dynamically. More specifically, we do our
calculations for the experimental parameters of Claussen {\it et al.}
\cite{neil}. In these experiments, $^{85}$Rb atoms are confined in a
cigar-shaped trap with radial frequency $\omega_r/2 \pi=17.4$ Hz and axial
frequency $\omega_z/2 \pi=6.8$ Hz, and the Feshbach resonance at $156.9$
(G)auss is used to vary the scattering length very rapidly. This is
achieved by a trapezoidal pulse in the magnetic field, which means that
the field  is ramped linearly to a certain value in a time  $t_{\rm
rise}$, then held for a time $t_{\rm hold}$, before ramping back to the
initial value. The initial and final values of the magnetic field are
chosen such that the scattering length is initially equal to zero, and is
large and positive during the hold. The rise time and the hold time are
typically of the order of microseconds, and therefore the shape of the
condensate wave function hardly changes during the pulse, but remains the
ground state of the trap, i.e., a gaussian.  However, the phase of the
condensate wave function changes considerably, and is in a gaussian
approximation given by
\cite{rembert1}
\[
 \theta_0 (t) + \frac{m}{2 \hbar} \sum_j \frac{x_j^2}{q_j (t)} \frac{d
 q_j(t)}{dt},
\]
with $q_j (t)$ the width of the gaussian in the three spatial directions.
Including both global and local phases of the condensate wave function in
our calculations is important, since, roughly speaking, these phases
determine the energy of a condensate atom. The variational parameters $q_j
(t)$ turn out to obey Newton's equations of motion \cite{collapse2}. The
equation of motion for the global phase $\theta_0 (t)$ is determined by
the condensate energy, which includes the effects of the evaporation
process.

For the retarded propagator of the ejected atoms, we use the expansion
\begin{eqnarray}
\label{expansionprop}
  G^{(+)} (\bx,t;\bx',t')&=& - i \theta (t-t') \sum_{\bn,\bm \neq 0} a_{\bn
  ,\bm} (t,t') \nonumber \\ && \times \chi_{\bn} (\bx) \chi^*_{\bm} (\bx') 
  e^{-\frac{i}{\hbar}(\epsilon_{\bn} t - \epsilon_{\bm} t')},
\end{eqnarray}
in which the sum is over all eigenstates, except for the ground state,
which is already contained in $\phi \args$. The coupled equations for the
expansion coefficients $a_{\bn,\bm} (t,t')$ are found from the equation of
motion for the Green's function in Eq.~(\ref{defgreens}). These
coefficients clearly obey the initial condition $a_{\bn,\bm}
(t,t)=\delta_{\bn,\bm}$ at the initial time, since all the atoms are then
in the condensate. 
\begin{figure}
\psfig{figure=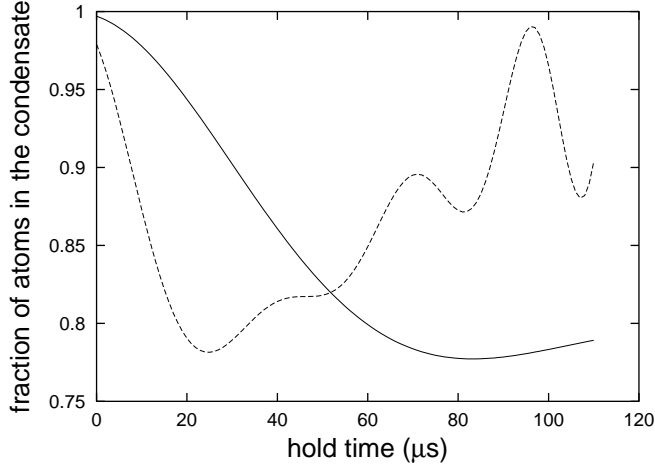} \caption{\narrowtext
	 The fraction of atoms in the condensate as a function of the hold
	 time, for a fixed rise time of $t_{\rm rise}=12.5 \mu s$. The
	 solid line corresponds to initially $N_{\rm c} (0)=6100$ atoms in
	 the condensate, whereas the dashed line corresponds to $N_{\rm c}
	 (0) =16500$  atoms. The scattering length during the hold is
	 equal to $a=2000 a_0$.
         \label{fig1}}
\end{figure}
\begin{figure}
\psfig{figure=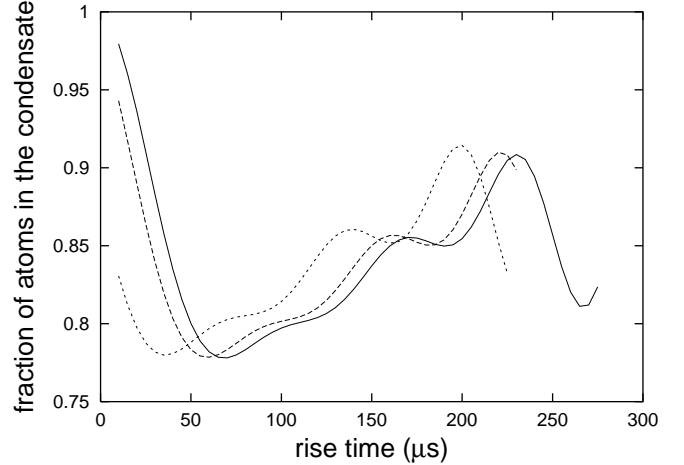} \caption{\narrowtext
         The fraction of atoms in the condensate, as a function of the rise
	 time, for different hold times, for a condensate of initially
	 $N_c(0)=16500$ atoms. The solid line corresponds to
	 $t_{\rm hold} = 1 \mu s$, the dashed line to $t_{\rm hold} = 5 \mu
	 s$, and the dotted line to $t_{\rm hold} = 15 \mu
	 s$. The scattering length during the hold is equal to $a=2000
	 a_0$.
         \label{fig2}}
\end{figure}

In Fig.~\ref{fig1} we show the fraction of atoms in the condensate as a
function of the hold time, for a fixed rise time of $t_{\rm rise}=12.5 \mu
s$. The initial number of condensate atoms is $N_{\rm c} (0) =16500$
atoms, and $N_{\rm c} (0)=6100$ atoms, respectively. The pulse is such
that the scattering length is equal to $a=2000 a_0$ during hold. Here,
$a_0$ is the Bohr radius. A significant fraction of the atoms is
transferred to the excited states in both cases, and part of this fraction can
come back into the condensate after some time, as seen from
Fig.~\ref{fig1}.  The atoms come back faster in the case of the largest
initial number of atoms, which  is caused by the fact that the coupling
between the condensate and the excited states is proportional to the
number of atoms. Note also that there are various frequencies in the
curve, which displays the fact that we are dealing with oscillations of
atoms between the condensate and several excited states. In
Fig.~\ref{fig2}, the fraction of atoms in the condensate as a function of
the rise time is displayed, for various hold times. Clearly, the number of
atoms increases with the rise time, over some range. This can, of course,
not be understood from the viewpoint of a loss mechanism characterized by
a rate constant, such as three-body recombination, or dipolar decay.

{\it Discussion.} ---
Comparing Fig.~\ref{fig1} to the experimental results of Claussen {\it et
al.} \cite{neil},  we notice that experimentally the number of atoms
always decreases with increasing hold time. An explanation for this
behavior is the presence of three-body recombination.  The comparison of
Fig.~\ref{fig2} with the experimental results leads to the same
conclusion.  The minima in this figure are seen to shift to the left, as
in the experiments, and also occur at the correct value of $t_{\rm
rise}$. However, experimentally they become lower with increasing hold
time, 
\begin{figure}
\psfig{figure=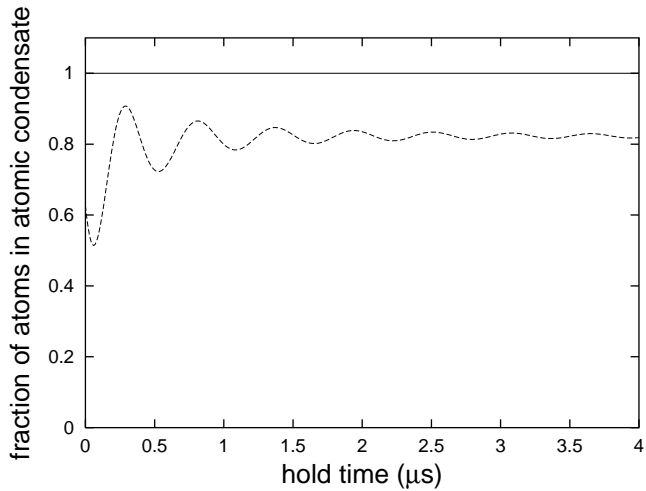} \caption{\narrowtext
	 The fraction of atoms in the condensate as a function of the hold
	 time, for a fixed rise time of $t_{\rm rise}=12.5 \mu s$. The solid
	 line corresponds to $a=2000 a_0$ and the dashed line to $a=4.9
	 \times 10^5 a_0$ during hold.
	 The initial number of condensate atoms is equal to $N_{\rm c} (0)=16500$ atoms.
	 \label{fig3} } 
\end{figure}
\noindent which can also be explained by a background loss mechanism
independent of the loss mechanism discussed here. 
Such an additional loss mechanism leads also to the smoothing of
the higher-order oscillations seen in our numerical results, which are not
observed in the experiments. Unfortunately, near a Feshbach resonance the
behavior of the three-body recombination is not known as a function of
the magnetic field. Therefore, it turns out that  a more detailed
comparison with experiment is impossible at this point.

A popular model for the description of a Feshbach resonance involves the
coupling of an atomic field to a molecular field
\cite{timmermans,murray}.   In such a model Rabi oscillations can also
occur. To the extend that we may neglect quantum evaporation, we can solve
this model essentially exactly using the framework developed here, by
making the substitution $T^{\rm 2B} |\phi|^2 \phi^* \hat \psi' \to g
(\phi^*)^2 \hat \psi_{\rm m}$ in $\hat H_{\rm int}$,  where $\hat
\psi_{\rm m}$ annihilates a molecule.  Note that we do not have to take
into account the anomalous averages of the atomic operators, since these
are effectively already incorporated by using the experimental values of
the coupling constants. Doing so would lead to a double counting of the
interaction effects \cite{nick}.  In the Heisenberg equation of motion for
the molecular field operator,  we have to incorporate the detuning from
the Feshbach resonance, defined as the energy difference between the
energy of the bound molecular state with respect to the atoms. It is given
by $\epsilon (t) = \kappa (B(t)-B_0)/\Delta B$, where $B_0$ and $\Delta B$
denote the position and the width of the resonance, respectively. The
prefactor $\kappa$ is fixed with the knowledge of the Zeeman effect of the
molecule with respect to the atoms \cite{servaas}, and the coupling $g$ is
subsequently determined to reproduce the correct strength of the Feshbach
resonance \cite{timmermans}. Since the $^{85}$Rb molecular bound state
responsible for the Feshbach resonance can not be trapped magnetically, we
perform our calculations for an optical trap, with  the same frequencies
as used in our previous calculations. In particular, this means that the
energy levels of the atoms and molecules are the same. We perform our
calculations for different hold times, but a fixed rise time of $t_{\rm
rise}=12.5 \mu s$. In Fig.~\ref{fig3} we present the result of our
calculations, for two different values of the magnetic field during hold
and for a condensate of initially $N_{\rm c} (0)=16500$ atoms. Note that
for the experimentally relevant case of $a=2000 a_0$  essentially no atoms
are converted to molecules. This can be understood from the fact that for
this value the detuning $\epsilon$ is much larger than the effective
coupling, i.e., $\epsilon \gg g \sqrt{n}$, with $n$ the density of the
atomic condensate. This means that the amplitude of the multimode Rabi
oscillations between the atomic condensate and the molecular gas,  which
is of order ${\mathcal O} \left( (g \sqrt{n}/\epsilon)^2 \right)$, is very
small. In order to convert a significant fraction of atoms to molecules,
we have to consider magnetic fields very close to the resonance. In
Fig.~\ref{fig3} we also show the results of our calculations for $a=4.9 \times
10^5 a_0$, which corresponds to a magnetic field that is $0.01$ G above
resonance. Due to the collective motion of the condensate the coupling
between the atomic and the molecular trap states decreases with time, which
leads to a damping and dephasing of the oscillations. Since we are dealing
with an enormous effective scattering length, the decay through three-body
recombination, and spin-flip processes becomes important. In our
calculations, we have however, for the same reasons as before,  not taken these
decay processes into account. They will lead to a further dephasing of the
Rabi oscillations.

In a future publication we intend to study the two pulse experiments
conducted recently \cite{elisabeth2}, and look at the properties of the
ejected atoms in more detail.

It is a pleasure to thank Servaas Kokkelmans, Neil Claussen, and Eric
Cornell for helpful remarks.

\enlargethispage*{2cm}

\end{multicols}
\end{document}